\documentclass{PoS}

\usepackage{amsmath,amssymb,amsthm,epsfig}

\newcommand{\beq}{\begin{equation}}
\newcommand{\eeq}{\end{equation}}
\newcommand{\bea}{\begin{eqnarray}}
\newcommand{\eea}{\end{eqnarray}}
\newcommand{\no}{\noindent}
\newcommand{\nn}{\nonumber}
\newcommand{\e}{\rm e\,}
\newcommand{\tr}{\hbox{tr}}
\newcommand{\Tr}{\hbox{Tr}}
\newcommand{\Det}{\hbox{Det}}

\newcommand{\one}{\hbox{1}}

\title{Discussion of the loop formula for the fermionic determinant}

\ShortTitle{On the loop formula}

\author{\speaker{Ion-Olimpiu Stamatescu}\thanks{Support from the Deutsche
Forschungsgemeinschaft for  attending the conference is thankfully acknowledged.}\\
        Inst. Theor. Physik, Univ. Heidelberg, Germany\\
        E-mail: \email{stamates@thphys.uni-heidelberg.de}}

\author{Erhard Seiler\\
        M.P.I. f. Physik, Munich, Germany\\
        E-mail: \email{ehs@mpp.mpg.de}}

\abstract{A formula expressing the lattice fermionic determinant 
(a large order polynomial) as an infinite product of 
smaller determinants is derived and discussed. These smaller determinants 
are of a fixed size, independent of the size of the lattice and are 
indexed by loops of increasing length.}

\FullConference{34th annual International Symposium on Lattice Field Theory\\
		24-30 July 2016\\
		University of Southampton, UK}

\begin{document}

\section{Introduction} \label{s.intro}

The study of the effects of virtual particles has a very long history. In 
particular the vacuum polarization due to electron-positron pairs was 
studied first by Euler and Heisenberg \cite{Heisenberg:1935qt} and later 
by Schwinger \cite{Schwinger:1951nm}. 

Among the later developments we can mention 
the vacuum polarization to 
all orders as given by the fermion determinant, whose properties were studied,
e.g., in \cite{seiler}, \cite{Fry:2015qua}. The  quantum effects on vortex
fields were analyzed by Langfeld et al 
\cite{Langfeld:2002vy},  Schmidt 
and Stamatescu \cite{Schmidt:2003bf}  pointed 
out that the fermion (and boson) determinant on the lattice can be viewed 
as a gas of closed loops and simulated numerically via a random 
walk - to quote only some of the more recent work.

Here we consider lattice gauge theories and  derive and 
discuss a general loop formula for the fermion determinant.
It is based on the loop expansion \cite{Stamatescu:1980br} 
and has been
used for HD-QCD \cite{dfss}, an approximation of QCD for large mass and 
chemical potential 
\cite{Bender:1992gn}, providing  
 systematic approximations to QCD. The formula 
can however lead to misinterpretations. We (re)derive it here explicitely
 and discuss its features in detail. See also \cite{loopfor}
 for a more general discussion.

\section{A simple example}

We consider 
\vspace{-0.5cm}{ \bea
\det\left(1-k(X+Y)\right)= \e^{ \tr\, \ln(1-k(X+Y))}
 \label{e.simple1}
\eea}
\no The traces  distinguish between
the strings $XXYY$ and $XYXY$, say, but
identify cyclic permutations, such as $XXYY$ and $XYYX$.
Expanding the exponent in (\ref{e.simple1})  we obtain:
\bea
&-&k\tr (X+Y)-\frac{1}{2}k^2\tr (X^2+2XY+Y^2)- ... 
-\frac{1}{4}k^4\tr (X^4+4X^3Y +2(XY)^2 +4X^2Y^2 +...) \nn\\
&=&-k\tr X-\frac{1}{2}k^2\tr X^2-\frac{1}{3}k^3\tr X^3 ... 
-k^2\tr XY-\frac{1}{2}k^4\tr (XY)^2 ...\label{e.simple2}
\eea
where we regrouped the terms observing the order
in which the monomials are formed in the products
$(X+Y)(X+Y)(X+Y)...$. 
We immediately see that 
resumming the terms which are  powers of a lowest order monomial
(what we shall call ``s-resummation'') 
we get the $\ln$ series.

To simplify the further discussion we shall now consider
 $X=x$ and $Y=y$ as just complex numbers, then:
{ \bea
 &&\ln(1-k(x+y)
 =\ln(1-kx)+\ln(1-ky)+\ln(1-k^2xy)...\label{e.simple4} \\
 &&1-k(x+y)= (1-kx)(1-ky)(1-k^2xy) 
 (1-k^3x^2y)(1-k^3xy^2)...\label{e.simple5}
 \eea}
  Obviously we have on the LHS of Eq. (\ref{e.simple5}) 
  a polynomial in $k$ while on the 
  RHS we have an
infinite product. 
Since the derivation is formally correct it is clear that 
 the validity of Eqs. (\ref{e.simple4}),(\ref{e.simple5}) 
 implies cancelations between
 infinite series which calls for convergence arguments.
 
  In particular in this example the
 LHS in Eq. (\ref{e.simple5})
 has just one zero at $k = \frac{1}{x+y}$ while the RHS appears to have
  an infinite
 series of zeroes at $k = 1/x, 1/y, 1/\sqrt{xy} , ...$. 
 For $k <\frac{1}{|x|+|y|}$
 convergence is ensured. The formula provides 
  a series of approximations of the LHS, 
 so, e.g. cutting after the 3-d order factor and expanding the product
 gives $1 -kx-ky + O(k^4)$,  correct to this order.
 
What we did was to apparently replace the one log cut on LHS
in (\ref{e.simple4}) by a superposition of log cuts on the RHS, correspondingly
the one zero of the LHS in (\ref{e.simple5}) by a superposition of
zeroes on the RHS. 
 The RHS zeroes (cuts) are not approximations of the LHS ones, 
but 
truncations of the product give approximations to the LHS which may be 
very good in the convergence domain.

For an illustration we may enquire which is
the variable's manifold on which the determinant vanishes. 
We find  
the zeroes of the RHS always lying  above the LHS one,   
with the lowest order ones (the straight lines)
 nearest to it. The first 3 factors give $1-\kappa (x+y) + \kappa^3
 x y (x+y)$, a 3-d order  approximation whose error increases drastically 
 outside the domain of convergence $x+y \leq 1/\kappa$.

\begin{figure}
\begin{center}
\epsfig{file=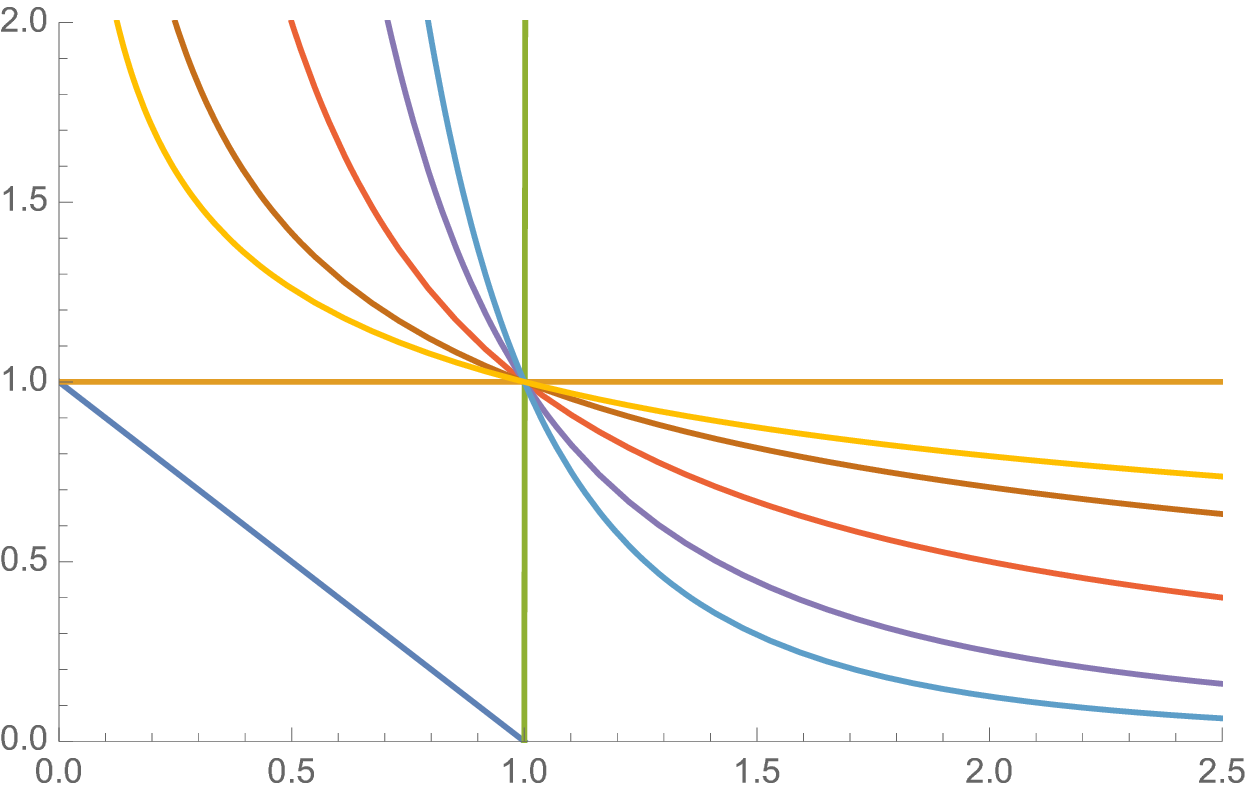, width=7cm}
\epsfig{file=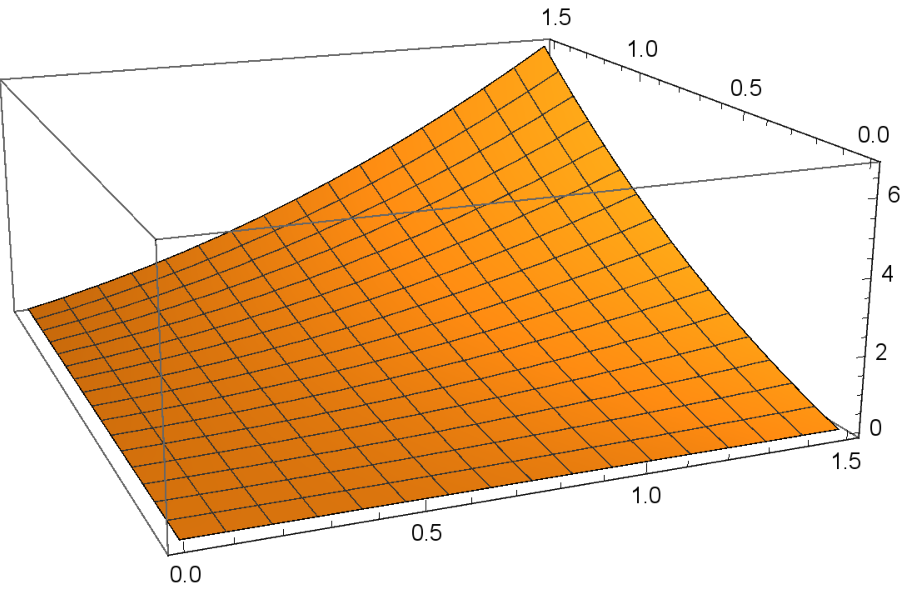, width=7cm}
\caption{{\em Left plot}: Zeroes of the RHS in 
the $x,\,y$ plane compared with the zero of the LHS (the diagonal
of the square). {\em Right plot}: The error of the 3-d order  approximation 
vs $x,\,y$. The unit is  $\frac{1}{\kappa}$.
}
\label{f.simple}
\end{center}
\end{figure}

\section{The loop formula for QCD with Wilson fermions in $d=2,\,4$.}

For definitness we give here the Wilson fermion matrix in $d=2,\,4$
for one flavour:
\bea
W &=&
\one
 -\kappa Q  \label{e.W}\\
&=&\one -  \kappa \sum_{i=1}^{d-1} \left( 
\Gamma_{+i}\,U_i\,T_i +
\Gamma_{-i}\,T^{-1}_i\,U^{-1}_i\right) 
 - \kappa\left( \e^{\mu}\,\Gamma_{+d}\,U_d\,T_d +
\e^{-\mu}\,\Gamma_{-d}\,T^{-1}_d\,U^{-1}_d \right)  \, \nn  \\
\Gamma_{\pm \nu} &= &1 \pm \gamma_{\nu},\, \gamma_{\nu}=\gamma_{\nu}^*,\,
\gamma_{\nu}^2 = 1,\,\tr \Gamma_{\pm \nu}= d ,
\label{e.Gamma}
\eea
($T$: lattice translations, $\kappa$: hopping
parameter, $\mu$ chemical potential). For latter generalisation we take
the links $U_{\nu} \in SL(3,C) \supset SU(3)$. 

The loop expansion and the loop formula for $\Det\, W$ are
\bea
\Det\, W &=& \Det( \one- \kappa Q) =
% {\rm exp} ( \Tr \ln (\one - \kappa Q) )     
 \e^{\Tr \,\ln (\one - \kappa Q) }      
  \label{e.hopg0} \\
 &=&  {\rm exp} \left[-\sum_{l=1}^\infty  \sum_{\left\{{\cal
C}_{l}\right\}} \sum_{s=1}^\infty ~{{{ g_{{\cal C}_{l}}}^s}\over
s}\,\tr_{\rm D,C}  \left[{\cal L}_{{\cal C}_{l}}^s\right] \right]  
\label{e.hopg1}\\
 &=& 
\prod_{l=1}^{\infty} ~\prod_{\left\{{\cal C}_{l}\right\}}  
  \det_{\rm D,C} \left(\one~-~ g_{{\cal C}_{l}}
{\cal L}_{{\cal C}_{l}}\right) \, \label{e.hopg2}\,, \\
g_{{\cal C}_{l}} &=& \kappa^{l}\,\left(\epsilon \, \e^{ N_{d}\mu}\right)^r, \ 
{\cal L}_{{\cal C}_{l}}= \prod_{\lambda \in {\cal C}_{l}}\Gamma_{\lambda}
U_{\lambda} \label{e.hopg3g}. 
\eea
Here ${\cal C}_{l}$ are {\em distinguishable, 
non-exactly-self-repeating lattice 
closed paths of length $l$} (called {\em primary paths} in the follwing).
$r$ is the net winding number of the path in the time direction ($d$),
with p.b.c. or a.p.b.c. and $\epsilon = +1(-1)$ correspondingly
( p.b.c. in the `spatial' directions). 
$\Det,\, \Tr$ imply Lattice, Dirac, and colour d.o.f.,
${\cal L}_{{\cal C}_{l}}^s \equiv \left({\cal L}_{{\cal C}_{l}}\right)^s$
 is the chain of links and $\Gamma$ factors along a {\em primary path}
(a {\em primary loop}), closing under the trace after $s$ repeted coverings of 
the path  ${\cal C}_{l}$. From Eq. (\ref{e.hopg1}) to (\ref{e.hopg2}) we
used ``s-resummation''.

{\em Derivation:} \\
$Q$ implies unit steps on the lattice and $Q^n$ 
generates a (closed) path 
 of length $n$,  with the weight $\frac{1}{n}$. This can be the $s$ 
 repetion of a closed path 
 of length $l$ - a {\em primary path}. 
 The primary path can start cyclically at
  each of its points, has therefore  multiplicity $l$, its repetitions
  do not bring new paths.
  % On even lattices $l$ is even.
  (NB: Pauli's 
  principle was 
  used obtaining the determinant, after that it's matrix algebra.)

The colour and Dirac traces close over the whole chain of length $l s$,
   the $s$ power of the {\em primary loop}  ${\cal L}_{{\cal C}_{l}}$
  Eq. (\ref{e.hopg3g}) and do not depend on the starting point 
  of the latter. Their contribution comes therefore with the weight 
  $\frac{l}{l s}= \frac{1}{s}$  and the
  factor  $g_{{\cal C}_{l}}$ coming from the links - see Eq. (\ref{e.hopg1}). 
We recognize here the logarithm series,
and partial
  summations over $s$ and exponentiation lead to Eq. (\ref{e.hopg2}).
  
  The loops in Eq. (\ref{e.hopg3g}) can be rewritten as
\bea
{\cal L}_{{\cal C}_{l}}&=& \Gamma_{{\cal C}_{l}} U_{{\cal C}_{l}},\
 \Gamma_{{\cal C}_{l}}=\prod_{\lambda \in {\cal C}_{l}}\Gamma_{\lambda},\  
U_{{\cal C}_{l}} = \prod_{\lambda \in {\cal C}_{l}} U_{\lambda} \ 
\label{e.lp1}\\
&&\tr_{D,C}{\cal L}_{{\cal C}_{l}}^s = \tr_D\Gamma_{{\cal C}_{l}}^s\,
\tr_C U_{{\cal C}_{l}}^s \equiv \tr\left(\Gamma_{{\cal C}_{l}}\right)^s\,
\tr\left( U_{{\cal C}_{l}}\right)^s \nn
\eea 
since the Dirac and colour traces factorise. The
Dirac factors $\tr\Gamma_{{\cal C}_l}^s \equiv \tr_D \Gamma_{{\cal C}_l}^s$ 
can be calculated for each ${\cal C}_l$ geometrically \cite{Stamatescu:1980br} 
or numerically. 

For loops exploring up to three dimensions we have moreover \cite{Stamatescu:1980br}
\bea
\frac{2}{d}\,\tr\left[\Gamma_{{\cal C}_{l}} \right]^s = 
\left[\frac{2}{d}\,\tr \Gamma_{{\cal C}_{l}} \right]^s = h_{{\cal C}_{l}}^s 
\label{e.lp5}
\eea
 which simplifies the contributions of these loops to
\bea
&&\det_C ( 1-g_{{\cal C}_{l}} h_{{\cal C}_{l}} U_{{\cal C}_{l}}) 
%\label{e.gh1}\\
=(1+C_{{\cal C}_{l}}^{3}) (1 + a\,\tr U_{{\cal C}_{l}} +
b\, \tr U_{{\cal C}_{l}}^{-1}),\label{e.gh2}\\ 
&&C_{{\cal C}_{l}}=g_{{\cal C}_{l}}h_{{\cal C}_{l}}, \
%a=  \frac{C_{{\cal C}_{l}}}{1+C_{{\cal C}_{l}}^{3}},\
a=  \frac{C_{{\cal C}_{l}}}{(1+C_{{\cal C}_{l}}^{3})},\ 
b= a\,C_{{\cal C}_{l}}
\label{e.gh3}
\eea

\section{Applications}

\subsection{HD-QCD}
For QCD at chemical potential $\mu > 0$ 
 the coefficients of primary loops of length $l$ with
 positive net winding number $r>0$ in the time direction are
\bea
g_{{\cal C}_{l}}  = 
\kappa^{l_{\sigma}} \epsilon^r \zeta^{rN_{\tau}},\,\ l_{\sigma}
= l-r\,N_{\tau} \geq 0 ,\,\  \zeta=\frac{d}{2}\kappa \e^{\mu} \label{e.zeta2}
\eea
Here $d=2,4$, $\epsilon =\mp 1$ for (a.)p.b.c..
Since $\zeta$ and $\kappa$ play different roles we order the
contributions according to $l_{\sigma}$.  

HD-QCD in leading order (LO, $l_{\sigma}=0$)
 ensues in the limit \cite{Bender:1992gn}
\bea
\kappa \rightarrow 0,\, \  \mu \rightarrow \infty , \, \ 
 \zeta: fixed \label{e.hadmlim}
\eea
It describes gluon dynamics with static quarks. 
Only the straight Polyakov loops $P$ in Eq. (\ref{e.hopg2})
survive. With $l_{\sigma}=2$ we retrieve Polyakov loops
with one decoration, $\tilde P$, and 
the quarks have some  mobility - \cite{dfss}. The
corresponding contributions are of the form Eq. (\ref{e.gh3}),
with 
\bea
C_P \equiv C = \epsilon (\frac{d}{2}\zeta)^{N_{\tau}},\ \
C_{\tilde P} \equiv C_r= \kappa^2 C^{r} ,
\eea
respectively. The decoration can be inserted anywhere along the
Polyakov loop and have any length. There are therefore
 $2(d-1)N_{\tau}(N_{\tau}-1)$
primary decorated loops of length $l = N_{\tau}+2$. From each of them 
we can form, however, 
further primary loops
of order $\kappa^2$ by attaching or inserting 
 any number $r>1$ of straight Polyakov loops to obtain .

We obtain to this order (up to a constant factor)
\bea
&&\det W^{[2]} =\prod_{\vec x}(1+a\, \tr P_{\vec x} +b\, \tr P_{\vec x}^{-1})\ 
\,\prod_{q}\prod_{r\geq 1}
(1+a_{r}\, \tr {\tilde P}_{q,r,\vec x} + b_{r}\, \tr {\tilde P}_{q,r,\vec x}^{-1}), 
\label{e.det_2}\\
&&a= C(1+C^3)^{-1},\ b =a\,C,\ \ a_{r} = C_r(1+C_r^3)^{-1},\ 
b_{r} =a_{r}\,C_r 
\eea
Here $q$ identifies the $2(d-1)N_{\tau}(N_{\tau}-1)$ shortest decorated 
Polyakov loops. The second factor in Eq. (\ref{e.det_2}), 
however, is an infinite product. 
For $\kappa$ small enough to ensure convergence 
we can cut the product, e.g. at $r=1$, this 
was done in \cite{dfss} for a reweighting simulation to
produce the phase diagram of QCD with 3 flavours of heavy quarks. 

\subsection{Complex Langevin Simulation}

The CL process associated to a partition function $Z$ with complex measure
proceeds
in the  manifold of   a complex variable $z$ and
involves a 
drift force $K$ as the logarithmic derivative of the 
measure
\bea
  Z= \int dz\, \rho(z),\ \ \delta z(t) =K(z)\delta t +\omega(z,t),\ \
 K(z) = \frac{\rho'(z)}{\rho(z)},  
\eea
with $\omega$ an appropriately normalized random noise.

A CL process to simulate QCD at nonzero $\mu$  takes place in the 
complexified space of the link variables $U \in SL(3,C)$ \cite{review},
\cite{Aarts:2014bwa}. 
The drift incorporates 
the logarithmic derivative of the determinant and needs the 
evaluation of the inverse of $W$, Eq. (\ref{e.W}) which is a large matrix
of rang $N_{\sigma}^{d-1} N_{\tau} N_c d$.

Using the loop formula Eq. (\ref{e.hopg2}) we have 
\bea
K(U_{\lambda})= \sum_{\left\{{\cal C}_{\lambda}\right\}} 
\frac{\partial_{U_{\lambda}}\det_{\rm D,C} \left(\one~-~ g_{{\cal C}_{\lambda}}
{\cal L}_{{\cal C}_{\lambda}}\right)}{\det_{\rm D,C} \left(\one~-~ 
g_{{\cal C}_{\lambda}}
{\cal L}_{{\cal C}_{\lambda}}\right)} \label{e.hoplang}
\eea
where the sum involves all loops ${\cal L}_{\lambda}$ 
which contain the link $U_{\lambda}$ and the  terms are easily calculable. 
There are of course infinitely many such loops and they may also contain 
powers of $U_{\lambda}$, a  simulation on these lines 
can only be achieved if we can meaningfully  limit the number of terms in
Eq. (\ref{e.hoplang}).

\section{Two more simple examples}

\begin{figure}
\begin{center}
\epsfig{file=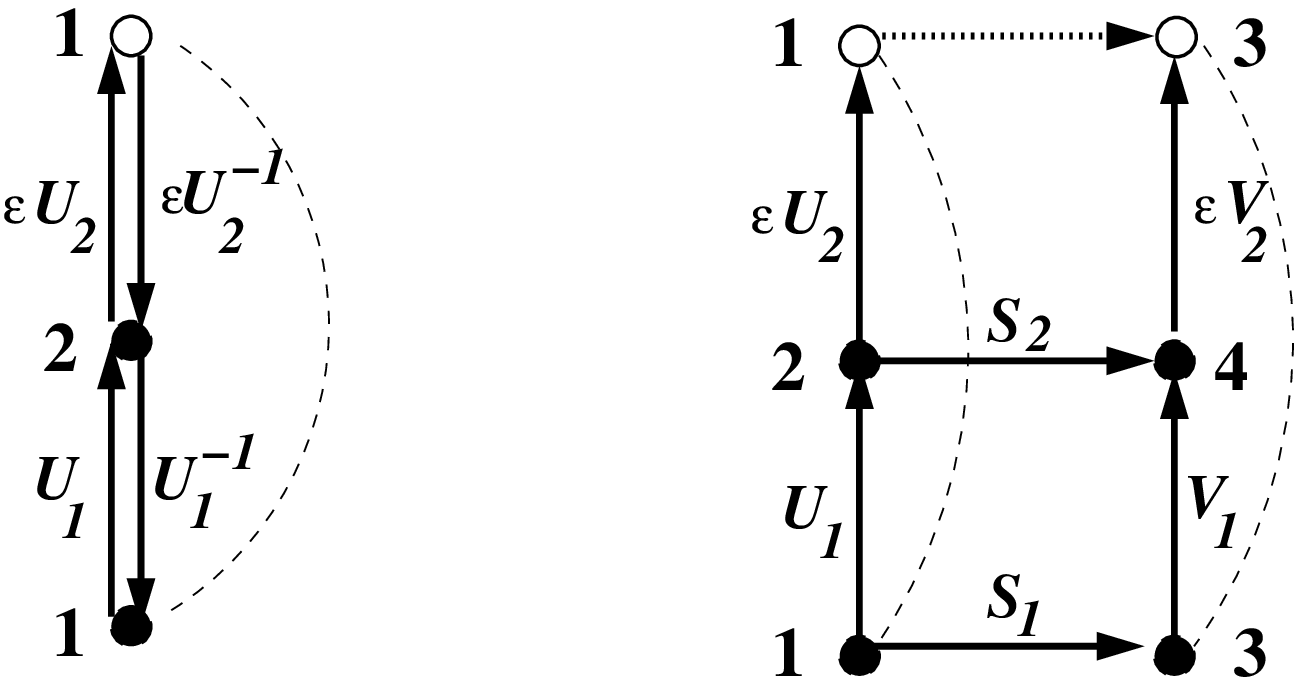, width=7.5cm}
\epsfig{file=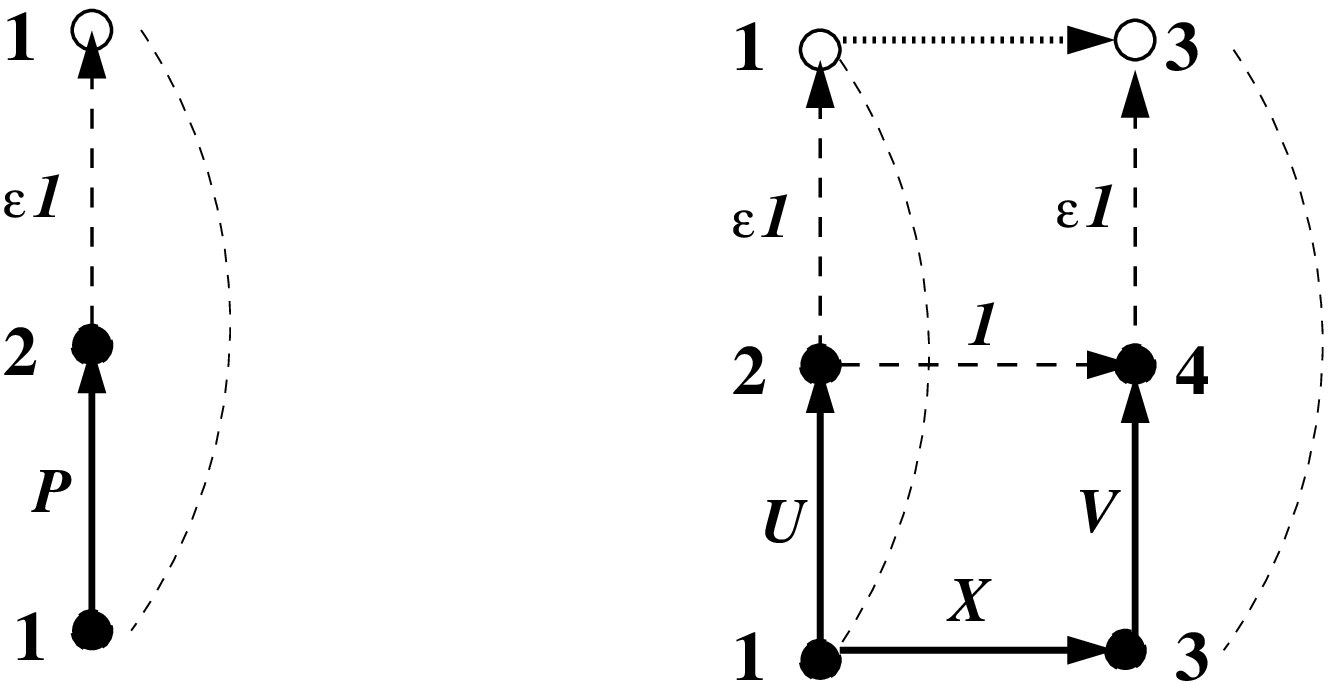, width=7.5cm}
\caption{Chain ({\em left} plot) and $2\times 2$ lattice 
without gauge fixing ({\em middle}) and in maximal gauge ({\em right} plot).
}
\label{f.examples}
\end{center}
\end{figure}

For illustration we present here 2 simple examples: a 1-d a.p.b.c. 
chain and a $2\times 2$ lattice with free b.c.in one direction and 
(a.)p.b.c. (for $\epsilon = \mp 1$, respectively)
in the other direction - see Fig. \ref{f.examples}.

The chain has only one primary loop  $P = U_1\,U_2$ and
the loop formula reproduces the exact  $\det W$
\bea
\det W =
 1+ 4\zeta^2 P +4\kappa^4 \zeta^{-2} P^{-1} +16 \kappa^4
\eea
A similar result  holds also for Polyakov loops of any length with a correspondig
power of $ \zeta^2$.

In the $2\times 2$ lattice  example there are 12 primary loops of length $l \leq 6$,  
listed here with their weights \footnote{The authors are indebted to E. Bittner
for providing a program to obtain primary loops.} 

\begin{align}
 &L_{1,2}:   && U_1 U_2 ,\   V_1 V_2 ,	&& (4\epsilon \zeta^2), \\ 
 &L_{3,4}:   && S_1 V_1 S_2^{-1} U_1^{-1},\ S_2 V_2 S_1^{-1} U_2^{-1},\, 
					&&(-4 \kappa^4) , \\
 &L_{5,6}:   && S_1 V_1 S_2^{-1} U_2,\ S_2 V_2 S_1^{-1} U_1,\, 
 					&&(4 \epsilon  \zeta^2\kappa^2) ,\\
 &L_{7-9}:   && S_1 V_1 V_2V_1 S_2^{-1} U_2,\
  S_2 V_2 S_1^{-1}U_1 U_2 U_1,\
  S_1 V_1 V_2 V_1 S_2^{-1} U_2,\ 	&&( -16   \zeta^4\kappa^2) ,\\ 
 &L_{10-12}: && S_2 V_2 V_1 V_2 S_1^{-1} U_1 ,\
  S_1 V_1 V_2 S_1^{-1} U_1 U_2,\ 
  S_2 V_2 V_1 S_2^{-1} U_2 U_1,\ 	&&( -16   \zeta^4\kappa^2) 
\end{align}
									 
The loop formula keeping only the loops of length up to 6 gives to
2-nd order in $\kappa$ (in maximal gauge) 
\bea
&&D^{[0]}(1,2) = 1-4\epsilon \zeta^2  (U + V) +16 \zeta^4 U\,V D^{[2]}(3-6)=
-4 \epsilon \kappa^2 \zeta^2 (V\,X
+ U\,X^{-1}) \label{hd2_2_1} \\
&&D^{[2]}(7-12)= 1 
-16 \kappa^2 \zeta^4 (2\,U\,V + X\,V\,U + X^{-1}U\,V +X^{-1}U^2 +X\,V^2))
\eea
and  we obtain the determinant to order $\kappa^2$
including all loops up to length 6, 
in complete agreement with the exact determinant to this order
\bea
\det W^{[2,6]} 
 =1 - 4\epsilon \zeta^2 (U+V) + 16 \zeta^4 U\,V 
- 4\epsilon \zeta^2 \kappa^2 (V\,X+U\,X^{-1}) -32  \zeta^4 \kappa^2 U\,V  
\eea

\section{Discussion}

As appealing as the loop formula appears its use is involved.
 The formula does not allow an interpretation as 
``linear factors'' decomposition, but
provides a systematic approximation in  $\kappa$ 
approaching the true determinant in the convergence domain.
As we have seen in sect. \ref{s.intro}, leaving 
the latter will provide a rather poor approximation.

On the other hand we can see the various orders in the loop
expansion as models by themselves and may analyse their
properties. In that case we have however to introduce some 
way of limiting the number of loops at any given length,
e.g. in a random walk generation of such loops.

The usefulness  of the loop formula relies therefore on its 
correct interpretation and adequate use.\par\bigskip

 {\bf Acknowledgment:} Support from the Deutsche
Forschungsgemeinschaft in the frame of the project  STA 283/16-2 is 
thankfully acknowledged.

\end{document}